# Toward an Efficient Website Fingerprinting Defense

Marc Juarez[1], Mohsen Imani[2], Mike Perry[3], Claudia Diaz[1], and Matthew Wright[2]

[1] KU Leuven, ESAT/COSIC and iMinds, Leuven, Belgium,
name.surname@esat.kuleuven.be
[2] The University of Texas at Arlington, TX, USA,
mwright@cse.uta.edu,mohsen.imani@mavs.uta.edu
[3] The Tor Project, https://torproject.org
mikeperry@torproject.org

**Abstract.** Website Fingerprinting attacks enable a passive eavesdropper to recover the user's otherwise anonymized web browsing activity by matching the observed traffic with prerecorded web traffic templates. The defenses that have been proposed to counter these attacks are impractical for deployment in real-world systems due to their high cost in terms of added delay and bandwidth overhead. Further, these defenses have been designed to counter attacks that, despite their high success rates, have been criticized for assuming unrealistic attack conditions in the evaluation setting. In this paper, we propose a novel, lightweight defense based on Adaptive Padding that provides a sufficient level of security against website fingerprinting, particularly in realistic evaluation conditions. In a closed-world setting, this defense reduces the accuracy of the state-of-the-art attack from 91% to 20%, while introducing zero latency overhead and less than 60% bandwidth overhead. In an open-world, the attack precision is just 1% and drops further as the number of sites grows.

**Keywords:** privacy, anonymous communications, website fingerprinting

## 1 Introduction

Website Fingerprinting (WF) is a type of traffic analysis attack that allows an attacker to recover the browsing history of a client. The attacker collects a database of web traffic templates and matches the client's traffic with one of the templates. WF has been shown to be effective in a wide variety of scenarios ranging from HTTPS connections [15], SSH tunnels [9], one-hop proxies [10], VPNs [19] and even anonymous communication systems such as Tor [5].

The success of WF against Tor, one of the largest deployed systems for anonymously browsing the Web [20], is particularly problematic. Tor offers stronger security than one-hop proxies and it is meant to protect against attacks like WF that require only a local eavesdropper or a compromised guard node. However, recent WF attacks achieve more than 90% accuracy against Tor [5,24,23], thus breaking the anonymity properties that it aims to provide.

To counter these attacks, a broad range of defenses has been proposed. The key building block of most of these defenses is *link padding*. Link padding adds varying amounts of delays and dummy messages to the packet flows to conceal patterns in network traffic. Given that bandwidth and latency increases come at a cost to usability and deployability, these defenses must strive for a trade-off between security and performance overheads. Unfortunately, the state-of-the-art link-padding defenses are not acceptable for use in Tor: they increase latency, delaying page loads between *two* and *four* times and impose bandwidth overheads between 40% [4] and 350% [8] on average.

We note that any delays introduced by a defense are a concern for low-latency systems, as they have a direct impact on the usability of the system in interactive applications. Moderate bandwidth overheads may also impact the user experience but the load factor needs to increase substantially before being noticeable by users. Moreover, the Tor network has spare bandwidth on the ingress edge of the network, making it possible to afford a client-side defense that consumes a moderate amount of bandwidth. In this work, we thus explore the design space of effective link-padding defenses with minimal latency overhead and modest bandwidth overhead.

The contributions of the following sections are:

**An analysis of the suitability of WF defenses for deployment in Tor.** In Section 2, we define the threat model and give a background of existing attacks and defenses. Based on this literature review, we discuss the suitability of these defenses for an implementation in Tor.

**A lightweight defense against WF attacks.** We have adapted Adaptive Padding to combat WF in Tor and dubbed this new defense *Website Traffic Fingerprinting Protection with Adaptive Defense* (WTF-PAD). Section 3 gives its specification, and Section 4 presents an evaluation and a comparison of WTF-PAD with the existing WF defenses. We find that WTF-PAD is effective and has reasonable overheads for a system like Tor.

**An evaluation of the defense in realistic scenarios.** Prior work has shown that the accuracy of the WF attack decreases significantly when certain assumptions about the setting or user behavior do not hold [11], but to the best of our knowledge this is the first study that evaluates the effectiveness of a WF defense in these scenarios. In Section 5, we show the results for two realistic scenarios: (i) *open-world*, in which the attacker monitors a small set of web pages and, (ii) *multi-tab*, where the users browse the pages using multiple tabs. We show that for these scenarios, the defense substantially reduces the accuracy of the state-of-the-art WF attack.

## 2 Website Fingerprinting (WF)

Tor is an overlay network that routes connections through three-hop circuits using *onion routing* [7]. The onion routers encrypt the messages in layers so that neither the relays nor localized network eavesdroppers can know both the origin and the destination of a connection.

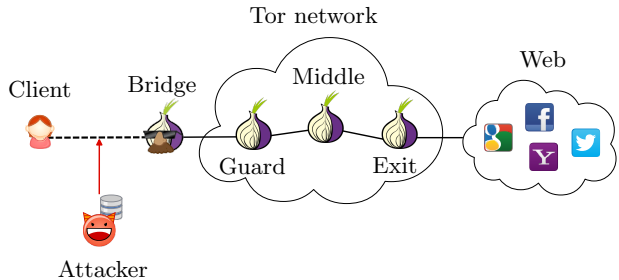

Fig. 1: The WF adversary model considering Tor bridges.

In this paper, we assume that the client connects to Tor through a *bridge*, a volunteer-run proxy to the Tor network (see Figure 1). The adversary has access to the communication at a point between the client and the bridge. The adversary is *local*, meaning that he is unable to observe other parts of the network, and *passive*: he can observe and record packets but cannot modify, delay, drop or inject new packets. We also assume that the adversary cannot learn anything about packet payloads due to the use of layered encryption.

Defensive padding is performed end-to-end between trusted endpoints, with the adversary only having access to the padded traces. For this research, we assume the bridge is trusted. This allows to implement the defense as a *Pluggable Transport* (PT) [21], avoiding modifications in the Tor source code. Note this model is equivalent for a client connecting to the trusted entry guard without a bridge, but in that case the defense would need to be implemented at the guard.

The objective of the WF adversary is to determine what pages the user downloads over Tor by looking at the network traces. Early works on this problem [10,19] assumed a user model that could only access a small set of pages—an assumption that is unlikely to be met in practice. This assumption is known as the *closed-world assumption*, and it overly simplifies the problem to the point of being irrelevant to most real-world settings. In contrast, the more realistic *open-world* allows the user to visit *any* page and the attacker's goal is to determine whether the user downloads one of a small set of *monitored* pages. We have evaluated both scenarios: the closed world favors the attacker and gives a lower bound of the defense effectiveness, but our objective is to measure the performance of the defense in realistic conditions.

WF attacks are a serious threat to Tor's security: the adversary only needs the ability to eavesdrop on the client's link to the network, which can be achieved with moderate resources. With the continuous improvement in WF classifier accuracy over the past few years, this is a pressing concern. The first attack against Tor obtained 3% accuracy with a Naive Bayes classifier [9] in a closed world and without any WF countermeasures. However, the attack has been revisited with more refined feature sets [17], and state-of-the-art attacks attain over 90% accuracy [5,24,23,16].

## 2.1 Defenses

Most of the defenses in the literature are theoretical designs without a specification for an implementation. Only a few have been evaluated for anonymous communications, and the only one that is currently implemented in Tor does not work as expected. In this section, we review WF defenses proposed in the literature and discuss their suitability for implementation in Tor.

**Application-level defenses.** These defenses work at the application layer. *HTTPOS* modifies HTTP headers and injects HTTP requests strategically [13], while *Randomized Pipelining*, a WF countermeasure currently implemented in the Tor Browser, randomizes the pipeline of HTTP requests. Both defenses have been shown to be ineffective in several evaluations [5,24,23,11].

**Supersequences and traffic morphing.** Recent works have proposed defenses based on generalizing web traffic traces [23,3]. They create anonymity sets by clustering pages and morphing them to look like the centroid of their cluster. This approach aims to optimally reduce the amount of padding needed to confound the attacker's classifier. These defenses, as well as traffic morphing techniques [25,12], have the shortcoming that require a database of webpage templates that needs to be frequently updated and would be costly to maintain [11].

**Constant-rate padding defenses.** Dyer et al. evaluated the impact of padding individual packets [8], finding that this is not sufficient to hide coarse-grained features such as *bursts* in traffic or the total size and load time of the page. Dyer et al. simulated a proof-of-concept countermeasure called *BuFLO*, which used constant-rate traffic with fixed-size packets. The authors report excessive bandwidth overheads in return for moderate security. The condition to stop the padding after the transmission ends is critical to adjust the trade-off between overheads and security. BuFLO stops when a page has finished loading and a minimum amount of time has passed, not covering the size of a page that lasts longer than the minimum time.

*Tamaraw* [4] and *CS-BuFLO* [5,2], both attempt to optimize the original design of BuFLO. Instead of setting a minimum duration of padding, Tamaraw stops padding when the total number of transmitted bytes is a multiple of a certain parameter. This approach groups webpages in anonymity sets, with the amount of padding generated being dependent on the webpage's total size. Given the asymmetry of web browsing traffic, Cai et al. also suggest treating incoming and outgoing traffic independently, using different packet sizes and padding at different rates. Furthermore, the authors sketched CS-BuFLO as a practical version of BuFLO, extended with congestion sensitivity and rate adaptation. Following Tamaraw's grouping in anonymity sets by page size, they propose either padding up to a power of two, or to a multiple of the amount of transmitted application data.

We question the viability of the BuFLO-based defenses for Tor. Their latency overheads are very high, such as two-to-three times longer than without defense, and the bandwidth overheads for BuFLO and CS-BuFLO are over 100%. In addition, due to the popularity of dynamic web content, it is challenging to determine when a page load completes, as needed in Tamaraw and CS-BuFLO. Nevertheless, in this paper, we compare our system against these defenses because they are the closest to meeting the deployment constraints of Tor.

## 3 Adaptive Padding

*Adaptive Padding* (AP) was proposed by Shmatikov and Wang to defend against end-to-end traffic analysis [18]. Even though WF attacks are significantly different from these end-to-end attacks, AP can be adapted to protecting against WF due to its generality and flexibility. AP has the defender examine the outgoing traffic pattern and generate dummy messages in a targeted manner to disrupt distinctive features of the patterns — "statistically unlikely" delays between packets. Shmatikov and Wang showed that with 50% bandwidth overhead, the accuracy of end-to-end timing-based traffic analysis is significantly degraded [18].

In the BuFLO family of defenses, the inter-arrival time between packets is fixed and application data is delayed, if needed, to fit the rigid schedule of constant packet timings. This adds delays in the common case that multiple real cells are sent all at once, making this family of defenses ill-suited for a system like Tor, as it would significantly harm user experience. By contrast, Adaptive Padding (AP) does not delay application data; rather, it sends it immediately. This minimal latency overhead makes AP a good candidate for Tor.

In the rest of this section, we describe AP and explain how we adapt it to defend against WF attacks in Tor.

### 3.1 Design Overview

To clarify the notation adopted in this paper, we use *outgoing* to refer to the direction from the PT instance running at the client to the PT at the bridge, and conversely, *incoming* is the direction from the PT server to the client.

The basic idea of AP is to match the gaps between data packets with a distribution of generic web traffic. If an unusually large gap is found in the current stream, AP adds padding in that gap to prevent long gaps from being a distinguishing feature. Shmatikov and Wang recognized the importance of bursts in web traffic and thus developed a dual-mode algorithm. In *burst mode*, the algorithm essentially assumes there is a burst of real data and consequently waits for a longer period before sending any padding. In *gap mode*, the algorithm assumes there is a gap between bursts and consequently aims to add a fake burst of padding with short delays between packets. In this paper, we follow Shmatikov and Wang and define a burst in terms of bandwidth: a burst is a sequence of packets that has been sent in a short time period. Conversely, a gap is a sequence of packets that are spread over a long timespan.

**AP algorithm.** The AP algorithm is defined by two histograms of delays that we call $H_B$ (used in burst mode) and $H_G$ (used in gap mode). The histograms have a set of bins that spans over the range of possible inter-arrival times. Each bin contains a number of *tokens*, which can be interpreted as the probability of selecting an inter-arrival time within the range of delays represented by that bin. The last bin, which we dub the "infinity bin", includes all possible values greater than the second-to-last bin. For more details on how these histograms are defined in WTF-PAD we refer the reader to Appendix A.

AP implements the state machine shown in Figure 2 in each defense endpoint, i.e. both PT client and server. For simplicity, let us consider just the client's state machine in the following explanation. The operation of the server is symmetrical.

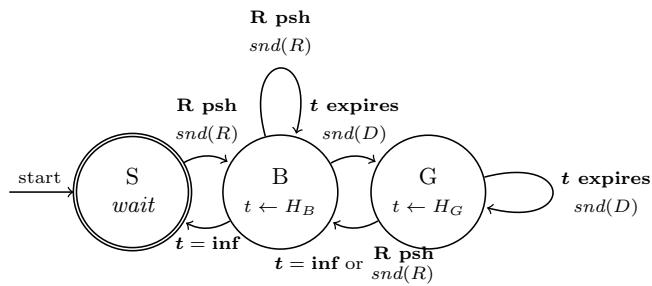

Fig. 2: AP algorithm as a finite state machine as implemented in the PT client. The events are in bold and the actions in italics. The action ($snd(\cdot)$) refers to sending messages, either *real* (R) or *dummy* (D). The **psh** event means a message pushed from the application (Tor browser) to the PT client.

**Burst mode.** As depicted in the diagram, AP starts idle (state $S$) until the packet with the HTTP request is pushed from the browser ($R$). This causes it to enter burst mode (state $B$), drawing a delay $t$ from the $H_B$ histogram. Then it starts to count down until either new data is pushed or $t$ expires. In the first case, the data is immediately forwarded, a new delay is sampled and the process is repeated again, i.e. it remains in burst mode. Otherwise, a dummy message ($D$) is sent to the other end and AP switches to state $G$ (gap mode). The $H_B$ histogram is built using a large dataset of web traffic, out of which we sample the times between the end of a burst and the beginning of the following burst (see Section 3.3). Therefore, while we are in a burst, the delays we sample from $H_B$ will not expire until we find an inter-arrival time that is longer than typical within a burst, which will make the delay expire and trigger the $G$ state.

**Gap mode.** While AP is in state $G$, it samples from histogram $H_G$ and sends dummy messages when the times it samples expire. The histogram for gap mode, $H_G$, is built from a sample of inter-arrival times *within* a burst in traffic collected for a large sample of sites. That is, by sending packets with inter-arrival times drawn from $H_G$, we are able to generate fake bursts that follow the timing distribution of an average burst. A transition from $G$ back to $B$ occurs upon either sampling a token from the infinity bin or receiving a real packet. Similarly, a transition from $B$ to $S$ happens when we sample a token from the infinity bin.

Note that AP immediately forwards all application data. Since sending a real packet means that the timeout expired, AP has to correct the distribution by returning the token to its bin and removing a token from the bin representing the actual delay. This prevents the combined distribution of padding and real traffic from skewing towards short values and allows AP to adapt to the current transmission rate [18]. If a bin runs out of tokens, to minimize its effect on the resulting distribution of inter-arrival times, we remove tokens from the next non-empty greater bin [18]. In case all bins are empty, we refill the histogram with the initial sample.

### 3.2 WTF-PAD

We propose a generalization of AP called *Website Traffic Fingerprinting Protection with Adaptive Defense (WTF-PAD)*. WTF-PAD includes implementation techniques for use in Tor and a number of link-padding primitives that enable more sophisticated padding strategies than the basic AP described above. These features include:

**Receive histograms.** A key feature to make padding realistic is to send padding messages as a response to messages received from the other end. In WTF-PAD, we implement this by keeping another AP state machine that reacts to messages received from the other PT endpoint: the PT client has a *rcv* event when it gets a packet from the PT server. This allows us to encode dependencies between incoming and outgoing bursts and to simulate request-response HTTP transactions with the web server. Padding introduced by the *rcv* event further distorts features on bursts, as just one packet in the outgoing direction might split an incoming burst as considered by the attacks in the literature.

**Control messages.** WTF-PAD implements control messages to command the PT server padding from the PT client. Using control messages, the client can send the distribution of the histograms to be used by the PT server. This way, the PT client is in full control of the padding scheme. It can do accounting on received padding traffic and alert the user if relays in its circuits are sending unscheduled padding.

**Beginning of transmission.** Control messages can also be used to signal the beginning of the transmission. If we are in state $S$ and a new page is requested, we will need to flag the server to start padding. Otherwise, the transmission from the first request to the following response is uncovered and reveals the size of the `index.html` page.

**Soft stopping condition.** In contrast to Tamaraw and CS-BuFLO, WTF-PAD does not require an explicit mechanism to conceal the total time of the transmission. At the end of the transmission, the padding is interrupted when we hit the infinity bin in the gap state and then the infinity bin in the burst state. See the Appendix A for further discussion on how to set the tokens in the infinity bins. The lack of a firm stop condition represents an advantage over existing link-padding-based defenses, which require a mechanism to flag the boundaries of the transmission. The probability of stopping will depend on the shape of the histograms at the end of the transmission.

### 3.3 Inter-arrival time distributions

Shmatikov and Wang did not specify in the original AP paper how to build and use the distribution of inter-arrival times in the AP histograms. In their simulations, they sampled the inter-arrival times for both real and padding traffic from the same distribution. To build the histograms, we have sampled the times from a crawl of the top 35K pages in the Alexa list. First, we uniformly selected a sample of approximately 4,000 pages and studied the distribution of inter-arrival times within their traces.

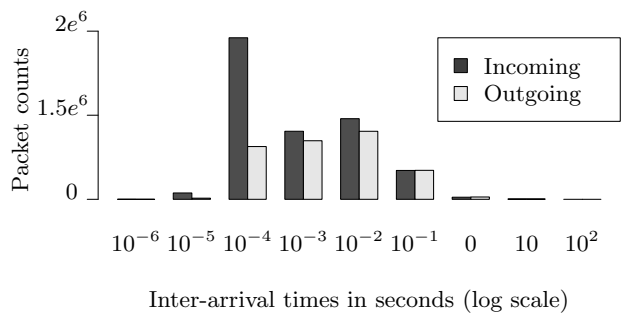

Fig. 3: Histogram of the inter-arrival times in a large sample of the top 35K Alexa.

In order to implement WTF-PAD without revealing distinguishing features between real and fake messages, we need to send dummies in time intervals that follow the same distribution as real messages. In Figure 3, we observe that times for incoming and outgoing traffic have different distributions. The asymmetric bit rates in the connection we used to conduct the crawl account for this difference. Since WTF-PAD has different histograms in the client and the bridge we can simulate traffic that follows different distributions depending on the direction.

Next, we explain how to find the bursts and the gaps in the inter-arrival time distribution and build the histograms $H_B$ and $H_G$. Intuitively, the burst-mode histogram $H_B$ should consist of larger delays covering the duration of typical bursts, while the gap-mode histogram $H_G$ should consist of smaller delays that can be used to mimic a burst. To split inter-arrival times into the two histograms, we calculate the instantaneous bandwidth at the time of each inter-arrival time to determine if it is part of a burst or not. Then, we set a threshold on the bandwidth to draw the line between bursts and gaps.

We estimate the instantaneous bandwidth using a sliding window over a sequence of consecutive packets. We have experimented with different window lengths and threshold values. The best results against the state-of-the-art WF attack are achieved for a window of two consecutive packets and a threshold set to the total average bandwidth for the whole sample of traces.

### 3.4 Tuning mechanism

AP can hide inter-arrival times that are longer than the average, but it does not hide times that are shorter than the average. To effectively hide these times we need to either add delays to exceptionally long traces or add more padding over all traces to level them off and make them less distinctive. We focus on the latter approach because our objective is to minimize delay. WTF-PAD provides a mechanism to tune the trade-off between bandwidth overhead and security: one can modify the parameters of the distributions used to build the histograms to add more padding and react to shorter inter-arrival times.

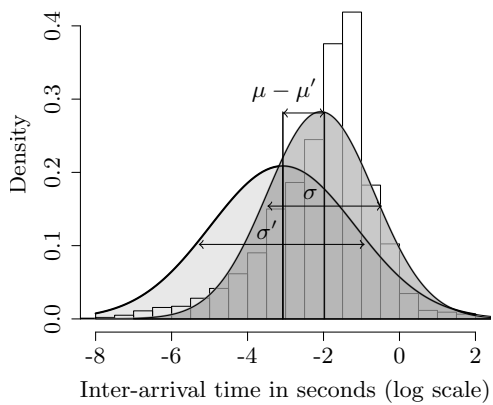

Fig. 4: Histogram of times between consecutive bursts for incoming traffic. In dark gray we superpose the PDF of our log-normal fit. In light gray, we show the PDF of a shifted log-normal distribution that we use to build the $H_B$ histogram.

To illustrate this, we show in Figure 4 the $H_B$ histogram as sampled from our dataset. We observe that the distribution of the logarithm of these times can be approximated with a normal distribution $\mathcal{N}(\mu, \sigma^2)$. That is, the inter-arrival times follow a log-normal distribution. We can modify its mean and variance to obtain another normal distribution $\mathcal{N}(\mu', \sigma'^2)$ that we will use to sample the inter-arrival times of $H_B$. By using $\mathcal{N}(\mu', \sigma'^2)$ we are shifting the average distribution of inter-arrival times toward shorter values. This results in a greater amount of short times being covered by padding, which increases the bandwidth overhead but causes the pages become less distinguishable and thereby reduces the attacker's accuracy.

We created a statistical model of the underlying distributions of inter-arrival times from the samples we extracted from our dataset. We experimented with multiple positively skewed distributions to build the model and test the goodness of fit with the Kolmogorov-Smirnov test. We estimated the parameters of the distributions using maximum likelihood estimation. Even though Pareto and Beta distributions seemed to fit best, we decided for simplicity to use normal and log-normal distributions, given that the error was not significantly greater than that observed in the other distributions.

To calibrate the possible shifts, we set $\mu'$ and $\sigma'$ according to the percentile of the real data we want to include. For instance, assuming a normal distribution, if we adjust $\mu'$ to the 50th percentile, we obtain $\mu' = \mu$ and $\sigma' = \sigma$. If we set $\mu'$ to the value of the Probability Density Function (PDF) at the 10th percentile, we then derive the $\sigma'$ using the formula of the PDF of the normal distribution.

## 4 Evaluation

In this section we discuss how we evaluated WTF-PAD, present our findings and compare them with the results we obtained for existing defenses.

### 4.1 Data

Unlike most previous defense evaluations, which used simulated data, we have used web traffic that has been collected over Tor. We used a dataset that had been collected for a study about a realistic evaluation of WF attacks [11]. This dataset consists of 40 instances, collected in ten batches of four visits, for each *homepage* in top-100 Alexa sites [1]. For the open-world, the dataset also has one instance for each website in the Alexa 35,000 most popular websites.

### 4.2 Methodology

To evaluate the improvements in performance offered by the defense, we applied the attack's classifier on both the original traffic traces and traces that have been protected by applying the defense. The difference in bandwidth and latency between the original and protected traces provides us with an estimate of the overheads. We applied the state-of-the-art attack on the set of protected traces to evaluate the effectiveness of the defense. The average accuracy over multiple runs determines the security provided by the defense.

In the closed world, we measure the accuracy as the True Positive Rate (TPR), or *Recall*. We also measure the False Positive Rate (FPR), the Positive Predictive Value (PPV)—also called *Precision*, and the harmonic mean of precision and recall (*F1-Score*), as they play an important role on evaluating the effectiveness of the attack in the open-world.

The state-of-the-art attack is based on a k-NN model [23]. k-NN is a supervised learning algorithm that selects the $k$ closest instances (the *neighbors*), and outputs the class of the majority of the neighbors. Wang et al. determined that the number of neighbors that optimizes the trade-off between TPR and FPR is $k = 5$. The distance defined by Wang et al. for use in k-NN is a weighted sum of a set of features. This feature set is the most extensive in the WF literature with more than 4,000 features and including features that extensively exploit bursts.

In order to have a comprehensive evaluation of WTF-PAD, we also evaluated it with other existing WF attacks that take into account features that are not included in k-NN.

### 4.3 Results

To evaluate the trade-off between bandwidth overhead and accuracy provided by WTF-PAD, we applied the attack on protected traces with different percentile values, ranging from 0.5 (low protection) to 0.01 (high protection) percentiles.

In Figure 5, we show the trade-off curves for both normal and log-normal fits. We observe a steeper decrease in accuracy for the normal model with respect to the log-normal one. Remarkably, beyond a certain point (around 0.1 percentile), the tuning mechanism saturates to 15% accuracy for both models: percentiles lower than that point do not further reduce accuracy and only increase bandwidth overhead. The trend we observe is the cost in bandwidth exponentially growing with the protection level that the defense attempts to provide

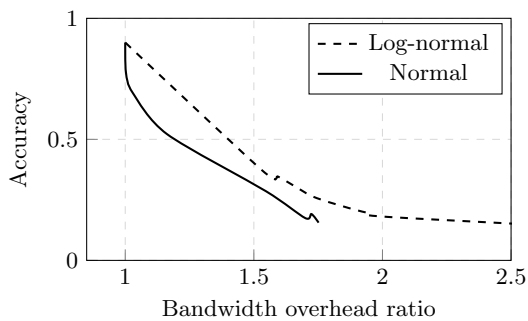

Fig. 5: Average accuracy versus median bandwidth overhead ratio.

Table 1 summarizes the security versus overhead trade-off obtained for different attacks (i.e., k-NN, NB, SVM, DL) and defenses BuFLO, Tamaraw, CS-BuFLO and WTF-PAD. As we see, WTF-PAD is the only defense to provide zero latency overhead. The other defenses we tested produce between 145-200% additional average delay to fetch a webpage. WTF-PAD also offers moderate bandwidth overhead. For our datasets, we observed that the bandwidth overhead was always below 60% while attaining decreases in the accuracy of the attack that are comparable with the other defenses.

Table 1: Performance and security comparison among link-padding defenses in a closed world.

| Defense | Parameters | Accuracy (%) | | | | Overhead (%) | |
|---|---|---|---|---|---|---|---|
| | | kNN | Pa-SVM [17] | DL-SVM [24] | VNG++ [8] | Latency | Bandwidth |
| BuFLO [8] | $\tau = 10s$, $\rho = 20ms$, $d = 1500B$ | 14.9 | 14.1 | 18.75 | N/A | 145 | 348 |
| CS-BuFLO [2] | $\rho = [20, 200]ms$, $d = 1500B$, CPSP | N/A | 30.6 | 40.5 | 22.5 | 173 | 130 |
| Tamaraw [23] | $\rho_{out} = 0.053$, $\rho_{in} = 0.138$, $d = 1500B$ | 13.6 | 10.59 | 18.60 | 12.1 | 200 | 38 |
| WTF-PAD | Normal fit, $p = 0.4$, $d = 1500B$ | 17.25 | 15.33 | 23 | 26 | 0 | 54 |

**ROC curve.** To study the impact of WTF-PAD on the performance of k-NN, we also plotted the ROC curve with and without protection. The ROC curve represents the performance of the classifier when its discrimination parameter changes. The standard k-NN is not a parametric algorithm, meaning that there is no explicit parameter that one can use to set the threshold and tune the trade-off. We have defined more or less restrictive classifications of k-NN by setting a minimum number of votes required to classify a page. We used 10-fold cross-validation to average the ROC curve for $k = 5$ neighbors in a closed world of 100 pages. To plot the ROC graph we had to *binarize* the classification: we divided the set of pages into two halves, 50 monitored and 50 non-monitored, and considered the monitored as the positive class and the non-monitored as the negative one. Then, all the positive (monitored) observations that are classified as a page in the positive class are counted as true positives, even if the instances were classified as a *different* monitored page. This is a more advantageous scenario for a surveillance-type of attacker that only tries to identify whether the page is monitored or not.

In Figure 6, we compare the ROC curves for the data before and after applying the defense with respect to random guessing. We notice a significant reduction in the performance of the classifier. Compared to unprotected data with an AUC of 0.95 (close to perfect classification), WTF-PAD has an AUC of 0.66, which is substantially closer to random guessing.

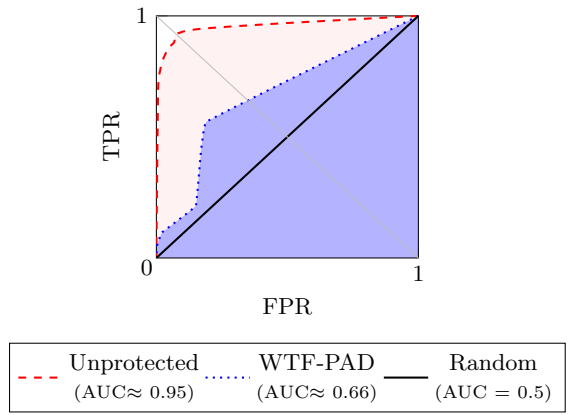

Fig. 6: 10-fold cross-validated ROC curves of k-NN with five neighbors and using a strict consensus threshold.

## 5 Realistic Scenarios

In this section, we present the results of the evaluation of the defense in two realistic scenarios: the open world and the use of multi-tab browsing.

### 5.1 Open-world evaluation

We now evaluate the performance of the defense against the k-NN algorithm in the open-world scenario. Our definition of the open-world is similar to the ones described in prior work. We have evaluated the k-NN with the evaluation method used by Wang et al. and incorporating the changes suggested by Wang [22], so that we can compare our results with the ones they obtained [23]

In Wang's open-world classification, they consider one class for each of the monitored pages and one single class for all the non-monitored pages. Then, the attacker aims to identify the exact monitored pages the user visits and to classify all the visits to non-monitored pages into the non-monitored class regardless of the actual page.

We observe that even though the accuracy initially increases as the world grows and saturates to 95% at the maximum considered world size, the F1-Score decreases and levels off to 50%. This is because even though the FPR rapidly drops to zero, the TPR decreases below 40%. The accuracy is so high because the classifier reaches almost perfect classification for the non-monitored class. This high accuracy is due to the stringent threshold used in the k-NN which requires all neighbors to vote to the same class and reduces the FPR.

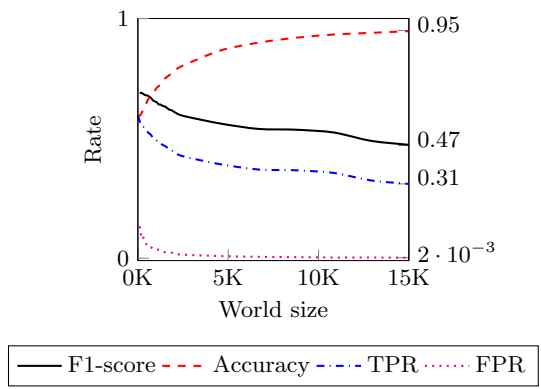

Fig. 7: Performance metrics for the classification of a k-NN classifier with $k = 4$ neighbors for an open world up to 15K pages [22].

We observe that the TPR and FPR after applying the defense are dramatically lower than the rates shown in Figure 7. However, due to the skew between the positive and the negative classes, the ROC curves of the k-NN are biased towards the negative class and do not reflect well the performance of the classifier. For imbalanced datasets, it is recommended to use the Precision-Recall ROC (P-ROC) instead of the ROC [6]. Similarly to the standard ROC, P-ROC represents the interaction of TPR (recall) and PPV (precision), instead of FPR, with respect to variations on the discriminant of the classifier. Precision in the open-world scenario conveys the fraction of monitored pages that were correctly detected by the k-NN. Precision is invariant to the size of the negative class and thus gives a more accurate estimation of the classifier's performance in the open-world.

In the P-ROC graph, the perfect classifier has a curve that coincides with the top-right corner and the random classifier is calculated as the number of positives divided by the total number of instances, i.e., the probability of selecting a positive instance uniformly at random. This random curve is used as a baseline because no classifier can have lower precision than it. As in the standard ROC, classifiers can be bench-marked by comparing their area under the curve (AUC).

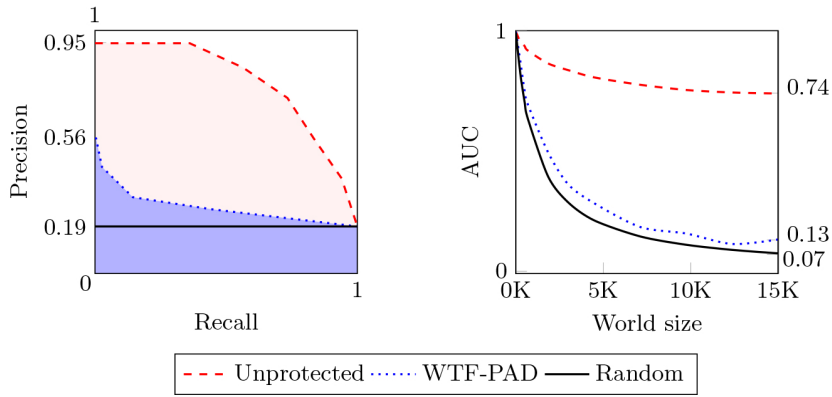

Fig. 8: The figure on the left shows the P-ROC curves for the k-NN attack on the protected and unprotected datasets for 5,000 pages. On the right, a comparison of P-ROC AUC with respect to the world size.

Figure 8 (left) shows the P-ROC curve of the k-NN when applied on the set of traces before and after WTF-PAD. Again, we observe that the AUC for the unprotected case is reduced significantly (from 0.79 to 0.27) and is close to random. However, this graph is a snapshot of the performance of the classifier for a fixed world size (5,000 pages). In order to evaluate how the size of the world affects the attack for the unprotected and protected data, we plot in Figure 8 (right) the AUC estimates while varying the size of the world. The first data point represents a closed world where all pages are monitored and, as expected, all classifiers perform as in perfect classification (AUC=1). However, as we increase the size of the world, the baseline classification tends to zero because a random guess is less likely to succeed. The k-NN levels off to AUC 0.74, which means that it is not heavily affected by the size of the world. Notably, when we apply the defense on the traces, all AUC values are close to random even for the largest world size that we have considered (15K pages). WTF-PAD steadily decreases the attack's success at the same rate as the random classifier does.

### 5.2 Multi-tab evaluation

The objective of the experiments in this section is to evaluate the efficacy of the WTF-PAD defense when the user is browsing with multiple tabs open. For this evaluation, we considered two scenarios and in both, the goal of the attacker is to identify one of the pages that compose the traffic trace.

In Scenario 1, we trained the k-NN attack on a single-tab dataset and tested on a mixed dataset of single tab traces and multi-tab traces generated by a crawl with two simultaneous tabs. The first tab was loaded following the Alexa top 100 sequentially. The second tab was open with a delay from 0.5 to 5 seconds and was chosen uniformly at random from the same list. Table 2 shows the result of Scenario 1 for traces with and without the protection offered by WTF-PAD.

Table 2: TPR for protected and unprotected traces in Scenarios 1 and 2.

|  | TPR | |
| --- | --- | --- |
|  | Unprotected | WTF-PAD |
| Scenario 1 | 14% | 8% |
| Scenario 2 | 68% | 22% |

Since the accuracy of the k-NN is already low when training on single-tab and testing on multi-tab (Scenario 1 in Table 2), the defense does not impact significantly the TPR of the classifier.

Table 3: TPR with respect to each traffic type. Each cell shows the number of background pages (the first tab) detected among truly detected multi-tab traces.

|  | Scenario 1 (TP/Total) | | | Scenario 2 (TP/Total) | | |
| --- | --- | --- | --- | --- | --- | --- |
|  | Single | Multi | First | Single | Multi | First |
| Unprotected | 233/300 | 901/8100 | 544/901 | 263/300 | 482/810 | 449/482 |
| WTF-PAD | 95/300 | 598/8100 | 333/598 | 108/300 | 137/810 | 103/137 |

In Scenario 2, we trained and tested k-NN on a dataset that includes multi-tab and single-tab traces. In this scenario, the attack achieves 68% TPR on unprotected multi-tab traces, much higher than the 14% found in Scenario 1. However, the success rate on protected traces drops to 22%.

In Table 3 we group the detection rates by traffic type (single or multi tab) as used to build the test set. k-NN can successfully classify unprotected single-tab traces with an accuracy of 87%, which is close to the accuracy rate of k-NN in the closed-world setting. The accuracy decreases to just 36% when we protect the traces with WTF-PAD.

## 6  Discussion and Future Work

WF attacks fall within the Tor threat model [7], as it only requires one point of observation between the client and the bridge or guard, and the attack potentially de-anonymizes users by linking them with their browsing activity. Even with the challenges of open-world and multi-tab browsing [11], some websites may exhibit especially unique traffic patterns and be prone to high-confidence attacks. Attacks may observe visits to the same site over multiple sessions and gain confidence in a result.

Protecting Tor users from WF attacks, however, must be done while maintaining the usability of Tor and limiting costs to Tor relay operators. Delay is already an issue in Tor, so adding additional delay would harm usability significantly. The BuFLO family of defenses add between 145-200% additional delay to the average website download, i.e. up to *three times as long* to get a webpage, which makes them very unlikely to be adopted in Tor.

The main overhead in WTF-PAD is bandwidth, which was under 60% overhead in all scenarios we tested. We do not know the exact percentage that is acceptable for use in Tor, but we note the following points. First, approximately 40% of Tor traffic is bulk downloads (from 2008, the last data we know of) [14]. To the extent that this holds today, only the remaining 60% of traffic needs to be covered by this defense. Second, the bottleneck in Tor bandwidth today is exit nodes. WF defenses do not need to extend to exit nodes, stopping at the bridge (in our framework) or at the guard or middle node when fully implemented. Thus, the bandwidth overhead only extends to one or two relays in a circuit and crucially not to the most loaded relay, making the overhead cost much less in practice. Third, given our findings for the open-world setting, it may be possible to tune WTF-PAD further to lower the bandwidth and maintain useful security gains in realistic use cases.

The construction of the histograms $H_B$ and $H_G$ is critical for the correct performance of the defense. First, since these distributions depend on the client's connection, we cannot estimate them a priori and ship them with WTF-PAD. A solution is to consider groups of clients with similar connections and have a precomputed configuration for each group. Then, the clients will estimate the properties of their network and only download the configuration that best matches their connection. Future work in developing WTF-PAD could explore the use of genetic algorithms to find the optimal histogram for each specific situation. A genetic algorithm could optimize a fitness function composed by the bandwidth overhead and the accuracy of the WF attack. Under mild assumptions on the distribution, histograms can be represented efficiently to reduce the search space.

## 7 Conclusion

In this paper, we described the design of WTF-PAD, a probabilistic link-padding defense based on Adaptive Padding. We studied the effectiveness and overheads of WTF-PAD, and compared it to existing link-padding-based defenses, showing that it offers reasonable protection with lower overhead costs. In particular, our results show that WTF-PAD does not introduce any delay in the communication while introducing moderate bandwidth overheads, which makes it especially suitable for low-latency communications such as Tor. Additionally, we have evaluated the effectiveness of WTF-AP in open-world and multi-tab scenarios. The results show that the defense reduces the performance of the classifier to random guessing.

## Acknowledgments

A special acknowledgement to Gunes Acar, Ero Balsa, Filipe Beato and Stijpan Picek for reviewing the draft version of the paper. We appreciate the interesting discussions with Yawning Angel, Rishab Nithyanand, Jamie Hayes, Giovanni Cherubin, Tao Wang, Oscar Reparaz and Iraklis Symeonidis that helped developing this paper. This material is based upon work supported by the National Science Foundation under Grants No. CNS-1423163 and CNS-0954133 and the European Commission through H2020-DS-2014-653497 PANORAMIX and H2020-ICT-2014-644371 WITDOM. Marc Juarez is funded by a PhD fellowship of the Fund for Scientific Research - Flanders (FWO).

## A  WTF-PAD Histograms

A histogram is defined as a disjoint partition of the support of the inter-arrival time distribution $[0, +\infty)$. Each sub-interval, that we call *bin*, is a half-closed interval $I_i = [a_i, b_i)$ with $0 \leq a_i, b_i \leq +\infty$ for all $i = 1, \ldots, n$, where $n \geq 2$ is the total number of bins in the partition. The bin lengths used in the AP histogram increase exponentially with the bin index, namely, the intermediate bins have the following endpoints:

$$a_i = \frac{M}{2^{n-i}}, b_i = \frac{M}{2^{n-i-1}},$$

for $i = 2, \ldots, n-1$. $M > 0$ is the maximum inter-arrival time considered in practice. The first bin is $I_1 = [0, \frac{M}{2^{n-2}})$ and the last bin is $I_n = [M, +\infty)$.

An exponential scale for the bins provides more resolution for values in a neighborhood of zero, which is convenient to represent distributions with heavy positive skew, such as the distribution of inter-arrival times in network traffic.

When we sample from a bin, AP returns a value sampled uniformly from $[a_i, b_i)$, except for the last bin $[M, +\infty)$, in which case AP returns "$\infty$".

In Figure 9, we show a simplified version of the histograms we used in the WTF-PAD instance at the client. The histograms that we use have 20 bins.

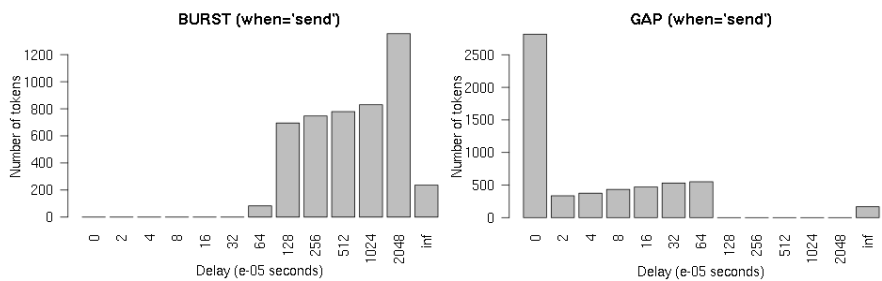

Fig. 9: Example of WTF-PAD histograms at the client. The histogram on the top is the $H_B$ and the one at the bottom is $H_G$.

Each bin contains a number of tokens $k_i$. We denote $K$ the sum of tokens in all the bins except the infinity bin, i.e.:

$$K := \sum_{i=1}^{n-1} k_i.$$

If we assume the probability of selecting a token is uniform over the total number of tokens, then the probability of sampling a delay from that bin can be estimated as:

$$P_i := \frac{k_i}{K + k_n}. \tag{1}$$

We assume that all the bins $I_i$ for $i < n$ are already filled as explained in Section 3.3. In the following we describe how to set the number of tokens in $I_n$, the infinity bin, for both histograms, $H_B$ and $H_G$.

**Infinity bin in $H_B$.** According to the notation introduced above, $P_n$ in $H_B$ is the probability of falling into the infinity bin and thus defines the probability of not sending padding (and not starting a fake burst) when we draw a sample from it. To express $k_n$ in terms of the probability of sampling from $I_n$ and the current sum of tokens in the histogram, we clear Equation 1 for $k_n$:

$$k_n = \frac{P_n}{1 - P_n} K.$$

For instance, if we decide on setting the probability of generating a fake burst to 0.9, then we need to set $P_n = 0.1$. Assuming $K = 300$ tokens, using the equation above we obtain $k_n \approx 34$.

**Infinity bin in $H_G$.** The number of tokens we will sample from $H_G$ until we hit the infinity bin is the number of dummy messages we will send within a fake burst. Since the probability of drawing a token is uniform, we can think the histogram as one single bucket that contains tokens from $I_n$ and tokens from the other bins. Then, the expected number of draws without replacement, $L$, until we draw the first token from the infinity bin is a known result one can find in any probability textbook:

$$E[L] = \frac{K + k_n + 1}{k_n + 1}.$$

We know the expected value of the length of a burst from our estimations on a large dataset of web traffic. Let $\mu_L$ be the mean burst length. In order to make sure fake bursts have the same mean length as real bursts, we must impose the expected number of tokens we sample until we hit the infinity bin to be: $E[L] = \mu_L$. Then, we only need to clear $k_n$ from the equation:

$$k_n = \frac{K - \mu_L + 1}{\mu_L - 1}.$$

This equation is well defined because, typically, the mean length of a burst is small: $K >> \mu_L$.